\documentclass[10pt,twocolumn,letterpaper]{article}

\usepackage{cvpr}
\usepackage{times}
\usepackage{epsfig}
\usepackage{graphicx}
\usepackage{amsmath}
\usepackage{amssymb}
\usepackage{algpseudocode}
\usepackage{algorithm}
\usepackage{algcompatible}
\usepackage{multicol}
\usepackage{multirow}
\usepackage{makecell}
\usepackage{hyperref}
\usepackage{bbold}
\usepackage{breqn}
\usepackage{tcolorbox}

\begin{document}

\title{Gumbel Rao Monte Carlo based Bi-Modal Neural Architecture Search for Audio-Visual Deepfake Detection}

\author{
Aravinda Reddy PN$^{1}$ \and
Raghavendra Ramachandra$^{2}$ \and
Krothapalli Sreenivasa Rao$^{2,1}$ \and
Pabitra Mitra$^{3,1}$ \and
Vinod Rathod$^{1}$  \and
\\
Indian Institute of Technology Kharagpur$^{1}$, India \\  
Norwegian University of Science and Technology$^{2}$, Norway. 
}


\maketitle
\thispagestyle{empty}

\begin{abstract}
Deepfakes pose a critical threat to biometric authentication systems by generating highly realistic synthetic media. Existing multimodal deepfake detectors often struggle to adapt to diverse data and rely on simple fusion methods. To address these challenges, we propose Gumbel-Rao Monte Carlo Bi-modal Neural Architecture Search (GRMC-BMNAS), a novel architecture search framework that employs Gumbel-Rao Monte Carlo sampling to optimize multimodal fusion. It refines the Straight through Gumbel Softmax (STGS) method by reducing variance with Rao-Blackwellization, stabilizing network training. Using a two-level search approach, the framework optimizes the network architecture, parameters, and performance. Crucial features are efficiently identified from backbone networks, while within the cell structure, a weighted fusion operation integrates information from various sources. By varying parameters such as temperature and number of Monte carlo samples yields an architecture that maximizes classification performance and better generalisation capability. Experimental results on the FakeAVCeleb and SWAN-DF datasets demonstrate an impressive AUC percentage of 95.4\%, achieved with minimal model parameters. \url{https://github.com/Aravinda27/GRMC-BMNAS}

\end{abstract}

\section{Introduction}
\label{sec:intro}
The advancement of deep generative models \cite{korshunov2018deepfakes} has brought about highly convincing synthetic audio and visuals, posing significant security risks. These technologies can potentially circumvent biometric systems that depend on unique individual characteristics. For instance, visual deepfakes employ techniques to change facial features, simulate wrongful acts, and modify appearances. Moreover, the latest developments in deepfake technology have made it possible to replicate human voices in real-time \cite{chintha2020recurrent}. Techniques for cloning voices use neural networks to create speech that sounds strikingly similar to a specific person, which complicates the reliability of authentication systems and opens up possibilities for impersonating public figures and committing financial deception.

Neural Architecture Search (NAS) \cite{liu2018darts} identifies optimal neural network designs within a predefined architecture space. Recent multimodal NAS (MMNAS) \cite{yu2020deep} explores attention mechanisms but relies on static network topologies.  Recently \cite{yin2022bm} proposed a multimodal DNN architecture that combined feature-level fusion with individual feature selection using Softmax. The impact of varying sample numbers on medical image analysis has been explored using the Gumbel-Softmax distribution within the NAS framework \cite{chang2019differentiable}. Recently \cite{pn2024straight} used Straight through Gumbel-Softmax based estimator-based bimodal NAS for audio-visual fake detection with reduced model parameters. However, the STGS-BMNAS suffers from high variance introduced by the Gumbel noise often suffers from unstable training dynamics.

\begin{figure*}[!h]
  \centering
   \includegraphics[width=1\linewidth]{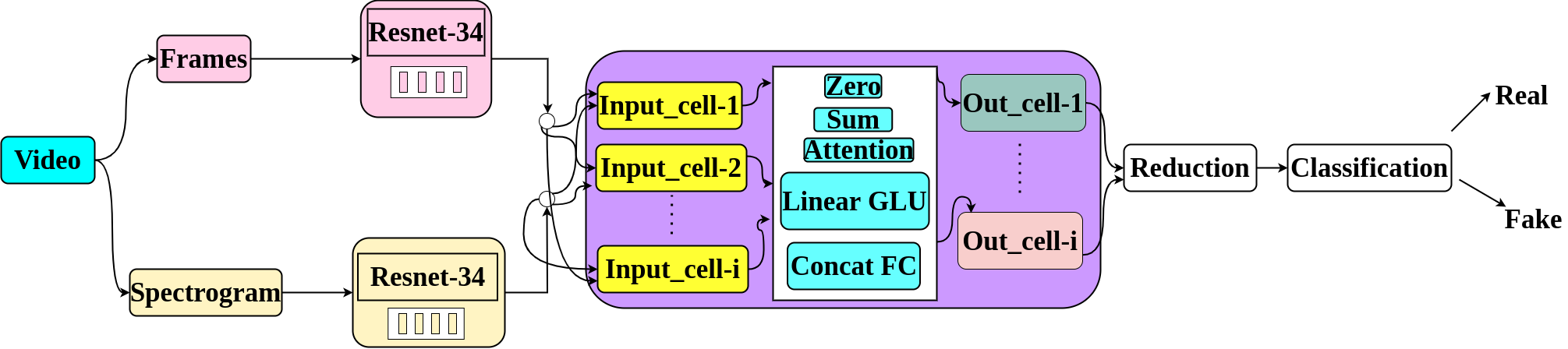}

   \caption{Block diagram showing the proposed GRMC-BMNAS employs a two-stage search to optimize bimodal fusion. The first stage identifies crucial features, while the second stage determines the optimal architecture using a pool of operations.}
   \label{fig:arch}
\end{figure*}

\begin{figure*}[!h]
  \centering
   \includegraphics[width=0.5\linewidth]{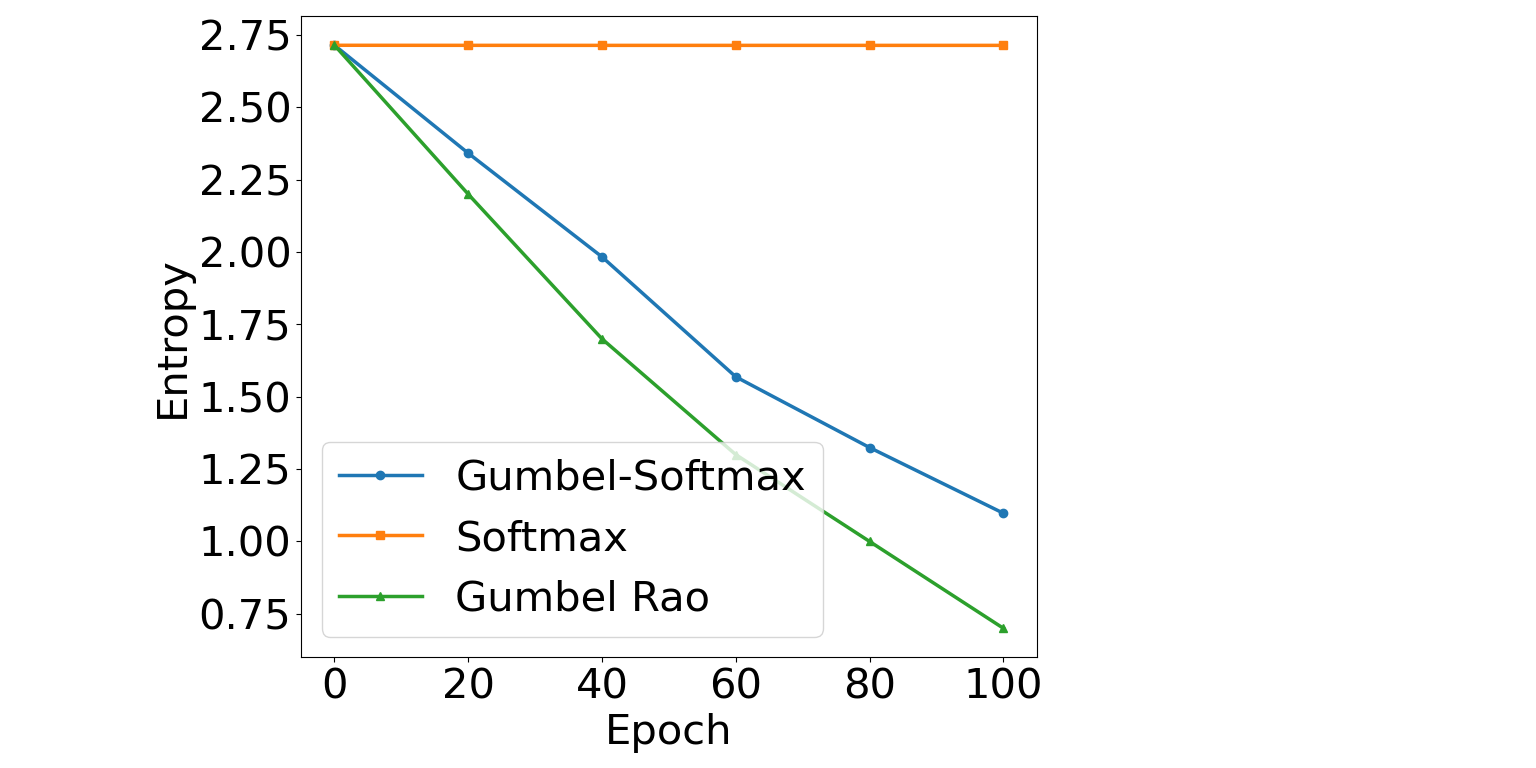}

   \caption{Average entropy plot for two learnable parameters for the proposed GRMC-BMNAS, and the existing STGS-BMNAS \cite{pn2024straight} and Softmax \cite{yin2022bm}.}
   \label{fig:entropy_arch}
\end{figure*}

The aim of this work is to develop a highly stable automatic architecture for audio-visual deepfake detection. So we propose Gumbel-Rao Monte carlo based bi-modal neural architecture search (GRMC-BMNAS) which adaptively learns architectures from a pool of operations for audio-visual deepfake detection and trains faster and offers better results on the test set performance. GRMC-BMNAS adopts a two-level search similar to \cite{pn2024straight} where it learns unimodal features from the backbone network by sampling the search space by varying the temperature parameter and Monte Carlo samples. In the second-level search, we utilize the weighted fusion strategy by varying the temperature and Monte Carlo samples. Increasing Monte Carlo samples expands the search space of primitive operations, allowing for a more accurate selection based on softmax probabilities as shown in Figure \ref{fig:arch}. As illustrated in Figure \ref{fig:entropy_arch}, the average entropy of GRMC-BMNAS consistently outperforms both STGS-BMNAS and the standard Softmax baseline \cite{yin2022bm}. This indicates that GRMC-BMNAS achieves a lower entropy, suggesting a faster convergence during training. Our proposed framework matches the performance of STGS-BMNAS \cite{pn2024straight} while requiring significantly less training time and computational resources (GPU days). Moreover, our model demonstrates superior generalization on test data. The main contributions of this paper are as follows:

\begin{itemize}
    \item To achieve faster, generalizable design of automatic architecture for bi-modal learning (extendable to multimodal learning), we propose an automatic approach named Gumbel-Rao Monte Carlo approximation based NAS for audio-visual deepfake detection which adopts two level schema.
\item The GRMC-BMNAS is an end-to-end framework which is fully searchable using two level schema. The Gumbel-Softmax trick uses Gumbel noise to approximate categorical samples in a differentiable manner. In GRMC, multiple Gumbel noise samples are drawn to enhance this approximation. Monte Carlo estimation averages these samples to approximate expectations, while Rao-Blackwellization conditions on discrete outcomes to reduce variance, improving the estimator’s efficiency.
\item Our study assessed the GRMC-BMNAS model for audio-visual deepfake detection through extensive experiments. Empirical evidence indicates that our model trains faster and has fewer parameters compared to existing state-of-the-art models.
\end{itemize}

The rest of the paper is organized as follows. Section 2 discusses about the related work, Section 3 presents proposed work, Section 4 discusses about datasets, experimental protocol, architecture search and evaluation and results, and Section 6 concludes the paper.

\section{Related work}

In recent years, deep learning techniques have been extensively used to create convincing fake videos by altering both visual and auditory components. Notable research includes methods like the Siamese Network proposed in Emotions Don’t Lie, \cite{mittal2020emotions} which compares affective cues from both modalities within a video. The Modality Dissonance Score (MDS) network, introduced in Not Made for Each Other \cite{chugh2020not}, highlights differences in audio-visual pairs using contrastive loss. Additionally, approaches focusing on phoneme-viseme \cite{agarwal2020detecting} mismatch reveal how deepfake techniques struggle to accurately replicate mouth shape dynamics corresponding to specific sounds. Techniques like multimodal trace extracts \cite{raza2023multimodaltrace} and cross-modal learning further enhance deepfake detection by analyzing audio-visual correspondences. Some methods even rely on self-supervised learning \cite{feng2023self} and anomaly detection using unlabeled real data. Overall, these advancements contribute to better identifying inconsistencies in deepfake videos. Recently \cite{oorloff2024avff} employs a two-stage approach. First, self-supervised learning extracts features from real videos. Subsequently, these features are fine-tuned for deepfake classification using supervised learning.

\section{Proposed method: GRMC-BMNAS}
\vspace{-0.2cm}
This work introduces a new framework called GRMC-BMNAS for deepfake detection, which optimizes network exploration by sampling from the Gumbel distribution and using the Monte Carlo approximation to average these samples. Rao-Blackwellized conditions on discrete outcomes reduce variance and improve estimator efficiency. At the first level, features from the backbones are sampled and cells are explored within a directed acyclic graph (DAG). A cell is a DAG consisting of ordered sequences on a node, where each node is a latent representation with directed edges linked to primitive operBMNASations that transform the node. The second level involves a DAG of nodes within a cell, each representing an operation chosen from a predefined pool.

\subsection{Gumbel distribution}
The Gumbel distribution, also known as Type I within the generalized extreme value distributions, is tailored for modeling extreme events and anomalies. A ‘Gumbel’ random variable, which adheres to this distribution, is characterized by a duo of parameters: location parameter $\mu \in \mathbb{R}$ and non-negative scale parameter $\beta \in \mathbb{R}_{\geq 0}$. The corresponding probability density and cumulative density functions are given by: \newline

\begin{equation}
\centering
  f(x)=\frac{1}{\beta}e^-{\frac{x-\mu}{\beta}}e^{-e^{-\frac{x-\mu}{\beta}}}
    \label{eq:pdf}
\end{equation}

\begin{equation}
\centering
 F(x)=e^{-e^{-\frac{x-\mu}{\beta}}}
    \label{eq:cdf}
\end{equation}

\begin{equation}
\centering
 F^{-1}(u)=-\beta log(-log(u))+\mu
    \label{eq:cdf}
\end{equation}



The inverse cumulative density function (ICDF) is also called quantile function given by equation \ref{eq:cdf} and equation \ref{eq:cdf} is used in inverse transform sampling to transform sample from Uniform distribution $U(0,1)$ into a Gumbel via a double logarithmic relation.


\subsection{Gumbel-max trick}
The Gumbel-max technique is a strategy for drawing samples from a categorical random variable denoted by $I\sim Cat(\pi)$. It involves the addition of Gumbel-distributed noise, which is independent and identically distributed, to the log probabilities before normalization. More specifically $I=\mathop{argmax}_{i \in D}\:\{G^{(i)}+log \:\theta_i\}\sim Cat(\pi)$,
where $G^{(1)}, G^{(2)}, G^{(3)}... G^{(D)}$ are the i.i.d samples drawn from Gumbel distribution \ref{eq:cdf}.

\subsection{Gumbel-Softmax distribution}
Instead of producing discrete or ‘hard’ samples from a categorical distribution that lacks structure, one can create ‘soft’ samples, which are especially beneficial for estimating gradients. To grasp the relationship between these hard and soft samples, it’s essential to analyze the hard samples when they are expressed in their one-hot encoded form, that is, 
$\mathbb{1}_\omega \in \{0,1\}^N$. Then $ z=onehot(\mathop{argmax}_{i \in D}\:\{G^{(i)}+log \:\theta_i\})$
 From \cite{jang2016categorical, maddison2016concrete} we derive PDF of this distribution and denoted by $GS(\pi,\lambda).$ More specifically, the $i^{th}$ index of soft sample $S_\lambda \: \in \{\mathbb{R}^N_{\geq 0}:|S_\lambda|=1\}$ is as defined in the Equation \ref{eq:act_gumbel}.  


\begin{equation}
  S_{i;\lambda}=\frac{exp((log \: \theta_i+\: G^{(i)})/\lambda)}{\sum_{j \: \in D}exp((log \: \theta_i)+G^{(j)})/\lambda)}
    \label{eq:act_gumbel}
\end{equation}
The temperature parameter $\lambda$ in the Gumbel-Softmax distribution modulates its entropy and that of its samples. It serves as a measure of how much the soft sample $S_\lambda$ deviates from a sample from $Cat(\pi)$. As $\lambda$ trends towards zero, the distribution samples shift towards one-hot representations, aligning the Gumbel-Softmax distribution closely with the categorical distribution.

Continuous one-hot vector relaxation excels in learning representations and sequences. However, for tasks requiring discrete values, such as reinforcement learning actions, compressed data, or architecture search, we discretize the continuous output using argmax. In this family, the forward computation of f is unchanged, but
backpropagation is computed “through” a surrogate. This surrogate is known as straight through gumbel softmax (STGS) defined in equation \ref{eq:stgs}.

\begin{equation}
  \nabla_{STGS}:=\frac{\partial\:f(S_\lambda)}{\partial\:S_\lambda}\:\frac{\partial\:S_\lambda}{\partial\:\phi}
  \label{eq:stgs}
\end{equation}


where $S_\lambda=softmax(\phi+G)$ with $\phi=log \theta$ and $G$ is the i.i.d Gumbel variable. 

\subsection{Gumbel-Rao Monte Carlo Estimator (GRMC)}
The GR estimator aims to reduce the variance associated with GS-ST by introducing a local expectation as follows:

\begin{equation}
  \nabla_{GR} = \frac{\partial f(D)}{\partial D}E\left[ \frac{d softmax_\tau(\Theta+G)}{d\Theta}|D \right]
\end{equation}

\begin{equation}
  \theta_j + G_j|D = \begin{cases}
-log(E_j) + logZ(\theta)  &\text{if j=i}\\
-log(\frac{E_j}{exp(\theta_j)}+ \frac{E_i}{Z(\theta)}) &\text{o.w.}
\end{cases}
\end{equation}

However, this expectation is in multiple variables and is not analytically tractable. Hence \cite{paulus2020rao} used Monte Carlo integration with K samples from $G|D$. 
\begin{equation}
    \nabla_{GRMCK} = \frac{\partial f(D)}{\partial D}\left[ \frac{1}{K}\sum_{k=1}^{K} \frac{d softmax_\tau(\theta+G^k)}{d\theta} \right]
\end{equation}
where $G^k \sim \theta + G|D$ i.i.d. using the reparameterization and K is the number of similar distributions or sampling size. Note that the total derivative $d softmax_\tau (\theta + G^k)/d\theta $ is taken through both $\theta$ and $G^k$. 

\subsection{Analysing our proposed GRMC-BMNAS}
To better understand our proposed method, we establish three propositions:
\begin{enumerate}
    \item \textbf{Proposition 1:} For the two learnable parameters $\alpha $ (first level search) and $\gamma$ (second level search) the Gumbel-Rao Monte Carlo (GRMCK) estimator yields lower variance compared to the Straight-Through Gumbel-Softmax (STGS) estimator then:
    
        Let the estimators denoted by (6) and (7) be $\nabla_{STGS}$ and $\nabla_{GRMCK}$. Let $\nabla_{\alpha} = dE[f(D)]/d\alpha$, $\nabla_{\gamma} = dE[f(D)]/d\gamma$ represent the actual gradient we are attempting to estimate. Then for all values $K\ge 1$, we have,

\begin{equation}
   E\left[ \left\| \nabla_{GRMCK} - \nabla_\alpha  \right\|^2 \right] \le E\left[ \left\| \nabla_{STGS} - \nabla_\alpha  \right\|^2 \right]
\end{equation}

\begin{equation}
 E\left[ \left\| \nabla_{GRMCK} - \nabla_\gamma  \right\|^2 \right] \le E\left[ \left\| \nabla_{STGS} - \nabla_\gamma  \right\|^2 \right]
\end{equation}



    \item \textbf{Proposition 2:} Increasing the number of Monte Carlo samples K in the GRMCK estimator further reduces mean squared error compared to STGS.
    Let $\theta^K_{GRMCK}$ be the GRMCK estimator with K samples, $\theta_{STGS}$ be the STGS estimator, $\theta$ be the true gradient, $Var(\theta^K_{GRMCK})$ be the variance of GRMCK with K samples, $Var (\theta_{STGS})$ be the variance of STGS estimator. Then, the mean squared error (MSE) for both estimators can be expressed as follows:
    $MSE(\theta^K_{GRMCK})=\mathbb{E}(\theta^K_{GRMCK}-\theta^2)$, $MSE(\theta_{STGS})=\mathbb{E} [\theta_{STGS}-\theta^{2}]$, then by Jensen inequality we have 
    \begin{equation}
        \mathbb{E}[(\theta^K_{GRMCK}-\theta^2)]\leq Var(\theta^K_{GRMCK})
    \end{equation}

    \item \textbf{Proposition 3:} Let $\theta_{STGS}$ be the gradient estimator using the STGS estimator and $\theta^K_{GRMCK}$  be the gradient obtained using GRMC estimator with K samples. Let $\theta$ denote the true gradient. The asymptotic biases of two estimators are given by
    \begin{enumerate}
        \item \textbf{Asymptotic bias of STGS estimator:}\newline
            The STGS estimator generally retains a non-zero bias as the number of samples from gumbel distribution increases
        \item \textbf{Asymptotic bias of GRMC estimator:}\newline
            The GRMC estimator designed to reduce variance and improve the gradient estimation, has an aymptotic bias that approches zero as the number of Monte Carlo sample K increases. Formally,\newline            
\begin{equation}
\lim_{ K\to \infty}\mathbb{E}[\theta_{STGS}]-\theta \neq 0 
\end{equation}

\begin{equation}
  \lim_{ K\to \infty}\mathbb{E}[\theta_{STGS}]-\theta= 0 
\end{equation}

        \item \textbf{Comparative statement:} 
        \begin{equation}
            |\lim_{ K\to \infty}\mathbb{E}[\theta_{GRMC}]-\theta| <|\lim_{ K\to \infty}\mathbb{E}[\theta_{GRMC}]-\theta| 
        \end{equation}

    \end{enumerate}

\end{enumerate}

\subsection{Modality feature extraction}
Similar to \cite{pn2024straight} we employ pre-trained ResNet-34 models for both image (facial) and speech feature extraction. Instead of using the final output, we extract features from intermediate layers of the neural network to capture richer and more abstract representations of the input data. Instead of using the final output, we extract features from intermediate layers of the neural network to capture richer and more abstract representations of the input data.
\subsubsection{First level search: GRMC relaxation over the cells}
We extract single-modal features from pre-trained backbone networks for both image (I) and speech (S) cues. These extracted features are denoted as ${I^{i}}$ and  ${S^{i}}$ respectively. Then we formulate the first level nodes in a sequence. Then $\mathbb{F}=\{I^{(1)}, I^{(2)}...,I^{(N_A)},S^{(1)},S^{(2)}...,S^{(N_B)}... \\ Cell^{(1)}...,Cell^{(N)}\}$
Let $\mathbb{F}^a$, $\mathbb{F}^b$ be any two nodes from $\mathbb{F}$. Let $\alpha$ be the weight parameter connecting between $\mathbb{F}^{(a)}$, $\mathbb{F}^{(b)}$ then each edge is selected based on the unary operation. Let $\mathbb{O}^F$ be the set of candidate operations:\\

$\mathbb{O}^F=\left\{ 
  \begin{array}{ c l }
  \centering
    Identity(x)=x & \quad \textrm{selecting an edge} \\
    Zero(x)=0                 & \quad \textrm{discarding an edge}
  \end{array}
\right.$

where each operation refers to a function $o{.}$ to be applied on the $cell^{(a)}$ then by applying the gumbel rao.

\begin{equation}
  \bar{o}(a,b)_\lambda= \sum_{o \in O} \left[  \frac{1}{K}\sum_{k=1}^{K}\frac{exp(\alpha_o^{(a,b)}+G^k_{(a)}|D)}{\sum_{o^{'}\in O}exp(\alpha_o+G^k_{(b)}|D)}\right] o(x)
\end{equation}

\begin{equation}
  \nabla_{GRMCK}=\frac{\partial f(\bar{o}(a,b))}{\partial\bar{o}(a,b))}\;\frac{\partial\bar{o}(a,b)}{\partial\phi}
\end{equation}


where K = sampling size which influences the entropy of the Gumbel Rao distribution and $\phi=log(\alpha)$.

A cell is densely connected and receives input from all its predecessors $o^v=\sum_{u<v} \bar{o}^{(u,v)}(o^{(i)})$.
In the evaluation stage, since we want deterministic predictions, the probabilities obtained from the Gumbel Rao distribution can be directly used without the need for sampling or argmax operation as, $(a,b)=\alpha_o(a,b)$. Using softmax probabilities during evaluation provides deterministic predictions without relying on sampling. Softmax offers probabilistic interpretations and is more robust to noise compared to the deterministic argmax approach.

\subsubsection{Second level: Weighted fusion}
Following the approach in \cite{pn2024straight}, we employ the same predefined candidate operations. Each operation takes two input tensors $x,y$, and produces an output tensor $z$, all of which have dimensions $\mathbb{R}^{N\times C\times L}$. These operations are detailed in Table 1. The second level of GR-BMNAS optimizes a weighted fusion strategy within a cell structure. A cell is a directed acyclic graph composed of nodes representing latent representations and edges representing operations. The cell's architecture is defined by edge and operation configurations, while weight parameters are learned during optimization.\newline
\textbf{Weighted fusion strategy:} In this stage, the inner structure of $Cells^{(n)}$ is an ordered sequence of $\mathbb{C}_n$ then 
\begin{equation}
    \mathbb{C}_n=\{I,S, N^{(1)},....N^{(M)}\}
\end{equation}

A cell comprises three nodes: an input node $in_c^{(i)}$ and two intermediate nodes $c^{(j)}$, $c^{(l)}$. The input node processes the backbone network's output and generates two intermediate representations using a weighted fusion of candidate operations determined by the Gumbel-Rao Monte Carlo method.
\begin{eqnarray}
\tiny
    =\sum_{o^s \in \mathbb{O}^s} \left[ \frac{1}{K}\sum_{k=1}^{K}\frac{exp((\gamma^{(i)}+G^k_{(i)})/D)}{\sum_{o' \in \mathbb{O}^s}exp((\gamma_{\bar{o}^s}^{(i)}+G^k_{(i)})/D)} \right] \\ \times w_i(f(c^{(j)},c^{l}))
\end{eqnarray}
The weights, $\gamma$ determine the contribution of candidate operations. During evaluation, we directly use the probabilities from the Gumbel-Rao distribution for decision-making, eliminating the need for sampling or selecting the maximum value $ o^{(i)}=\gamma_c^{(i)}$. The edge weights $(\beta)$ are also relaxed using straight-through gumbel rao similar to the first level. The output node combines the results from all transformation nodes.

\begin{table*}[!t]
\centering
\begin{tabular}{c|c}

\hline
\textbf{Operation} & \textbf{Function}  \\ \hline
Zero(x,y) & \makecell{The Zero operation, eliminates an entire node,\\
effectively discarding its contribution}. \\ \hline
Sum(x, y): & \makecell{The DARTS  framework \cite{liu2018darts}, introduced , \\ employs a method to combine two features using summation. \\ 
    $Sum(x,y) = X+Y$
    }
    \\ \hline
Attention$(x,y)$ & \makecell{ The Attention operation, as described in \cite{vaswani2017attention}, employs scaled dot-product attention,\\  where a query $x$ and key-value pairs $y$ are used. \\
        $Attention(x,y)=Softmax(xy^T/(\sqrt{C}\times y))$
    } \\ \hline
 LinearGLU$(x,y)$& \makecell{The LinearGLU operation combines two inputs $x,y$,\\ using a linear layer followed by the gated linear unit (GLU) activation \cite{dauphin2017language}. \\
    $LinearGLU(x,y)=xW_1\bigodot Sigmoid(yW_2)$
   } \\ \hline
ConcatFC$(x,y)$ & \makecell{ The ConcatFC operation involves concatenating two inputs, \\
$(x,y)$ and and passing the concatenated vector through a fully connected (FC) layer with ReLU activation. \\ 
        $ConcatFC(x,y)=ReLU((x,y).W+b)$
    }
    \\ \hline
\end{tabular}
\caption{Candidate operations used in the second level search}
\label{tab:second_cand}
\end{table*}

\begin{table*}[!h]
\centering

\begin{tabular}{c|c|c|c|c|c}
\hline
\textbf{Method} & \textbf{AUC}(\%) &\textbf{ACC}(\%) &\textbf{Params(M)}  &\textbf{\makecell{GPU \\ days}}  &\textbf{Search}  \\ \hline
Voice-face \cite{cheng2023voice}&82  & 86 &174  & - &Gradient \\ 
\makecell{Audio-visual \\ anamoly detection \cite{feng2023self}}&93  & - &41  &- &Gradient \\ 
\makecell{Not made \\ for each other \cite{chugh2020not}} & 81 & 84.56 &122  &- &Gradient  \\
 ID-Reveal \cite{cozzolino2021id}& 78 & 80.1 &7.3  &- &Gradient \\ 
MultimodalTrace \cite{raza2023multimodaltrace} & 84 & 91.26 &15  &-  &Gradient\\ 
 Ensemble-learning \cite{hashmi2022multimodal}& 84 & 86 &12  &- & Gradient\\ 
POI-AV \cite{sung2023hearing} & 93.9 & 90.9&-  &- &- \\ 
BM-NAS \cite{yin2022bm} &92.26  & 91.4 &0.62  & 4&Gradient \\ 
     STGS-BMNAS\cite{pn2024straight}& 94.4 & 95.5 &0.26  & 2&\makecell{Straight\\ through estimator} \\ 
     \textbf{GRMC-BMNAS} (Ours) &\textbf{95.5}& \textbf{96.5}&\textbf{0.20}& 1.5&\textbf{\makecell{Straight\\ through estimator}}\\\hline
 
\end{tabular}
\caption{Comparison of our proposed GRMC-BMNAS with SOTA approaches tested on our test data}
\label{tab:SOTA_cmp}
\end{table*}

\begin{table*}[!h]
\centering

\begin{tabular}{c|cccccccc}
\hline
\multirow{4}{*}{\makecell{\textbf{Trained on}\\ $\downarrow$}} & \multicolumn{8}{c}{\makecell{\textbf{Tested on}\\$\downarrow$}}                                                                                                                                                    \\ \cline{2-9} 
                  & \multicolumn{4}{c|}{\textbf{GRMC-BMNAS}}                                                                         & \multicolumn{4}{c}{\textbf{STGS-BMNAS}}                                                    \\ \cline{2-9} 
                  & \multicolumn{2}{c|}{FakeAVCeleb}                         & \multicolumn{2}{c|}{SWAN-DF}                         & \multicolumn{2}{c|}{FakeAVCeleb}                         & \multicolumn{2}{c}{SWAN-DF}    \\ \cline{2-9} 
                  & \multicolumn{1}{c|}{AUC} & \multicolumn{1}{c|}{ACC} & \multicolumn{1}{c|}{AUC} & \multicolumn{1}{c|}{ACC} & \multicolumn{1}{c|}{AUC} & \multicolumn{1}{c|}{ACC} & \multicolumn{1}{c|}{AUC} &ACC  \\ \hline
                  FakeAVCeleb& \multicolumn{1}{c|}{94.7} & \multicolumn{1}{c|}{93.5} & \multicolumn{1}{c|}{91.6} & \multicolumn{1}{c|}{91.2} & \multicolumn{1}{c|}{92.7} & \multicolumn{1}{c|}{91.8} & \multicolumn{1}{c|}{85.6} & 84.7 \\ \hline
                  SWAN-DF& \multicolumn{1}{c|}{90.8} & \multicolumn{1}{c|}{91.2} & \multicolumn{1}{c|}{95.1} & \multicolumn{1}{c|}{94.8} & \multicolumn{1}{c|}{84.8} & \multicolumn{1}{c|}{83.2} & \multicolumn{1}{c|}{93.1} & 92.8 \\ \hline
                  
\end{tabular}
\caption{Generalisation of our proposed model to the seen and unseen data}
\label{tab:gen_data}
\end{table*}

\begin{table*}[!h]

\centering

\centering
\begin{tabular}{cc|ccc}

\hline
\multicolumn{2}{c|}{\multirow{2}{*}{ \textbf{Temperature ($\lambda$)} }}    & \multicolumn{3}{c}{\textbf{No of Samples (K)}}                                                    \\ \cline{3-5}
\multicolumn{2}{c|}{}                     & \multicolumn{1}{c|}{\textbf{10}} & \multicolumn{1}{c|}{\textbf{100}} & \textbf{1000}  \\ \hline
\multicolumn{1}{c|}{\multirow{2}{*}{$\lambda=0.1$}} & AUC  & \multicolumn{1}{c|}{92.16} & \multicolumn{1}{c|}{95.5} & 96.96  \\  
\multicolumn{1}{c|}{}                  & \makecell{Model \\ parameters} & \multicolumn{1}{c|}{341760} & \multicolumn{1}{c|}{205565} & 189574  \\ \hline
\multicolumn{1}{c|}{\multirow{2}{*}{$\lambda=0.5$}} &AUC  & \multicolumn{1}{c|}{91.75} & \multicolumn{1}{c|}{94.04} & 90.45   \\  
\multicolumn{1}{c|}{}                  & \makecell{Model \\ parameters} & \multicolumn{1}{c|}{322456} & \multicolumn{1}{c|}{192452} & 175642  \\ \hline
\multicolumn{1}{c|}{\multirow{2}{*}{$\lambda=1.0$}} & AUC & \multicolumn{1}{c|}{91.16} & \multicolumn{1}{c|}{93.95} & 90.2   \\  
\multicolumn{1}{c|}{}                  & \makecell{Model \\ parameters} & \multicolumn{1}{c|}{299490} & \multicolumn{1}{c|}{187852} & 167845  \\ \hline

\end{tabular}
\caption{Evaluation of searched architecture with different temperature values and with varying Monte carlo samples}
\label{tab:ablation_temp_sample}
\end{table*}

\begin{figure*}[!h]
\centering
\includegraphics[width=1\linewidth]{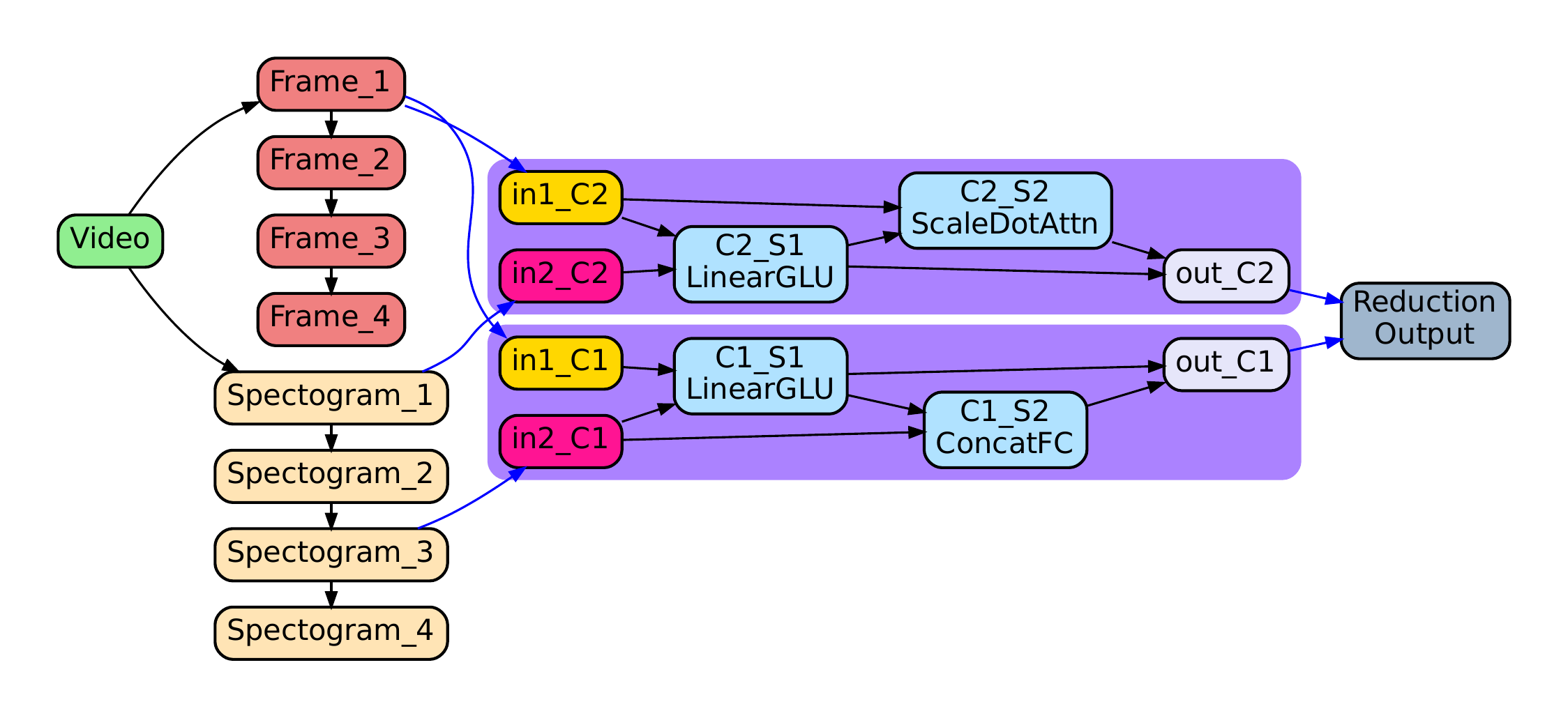}

   \caption{Best architecture obtained with K=100 and $\lambda=0.1$}
   \label{fig:optimal_arch}
\end{figure*}

\subsection{Optimizing Neural Architectures through Parameter learning}
We employ our proposed GRMC method to jointly optimize both weight parameters and architecture in an end-to-end training process. The objective of architecture search is to minimize the loss function, $\mathbb{L}_\omega$ while simultaneously reducing the number of model parameters i.e., $\underset{\omega,\alpha,\gamma}{min} \mathbb{E}_{\mathbb{A}\sim p_{(\alpha,\gamma)_{\lambda, K}}}|\mathbb{L}_\omega(\mathbb{A}))|$. The primary objective is to minimize the expected performance of architectures sampled from the search space i,e., $p_{(\alpha,\beta,\gamma)_{\lambda,K}}(\mathbb{A})$. Our method involves sampling network architectures from a distribution parameterized by $\alpha$, $\beta$, and $\gamma$, controlled by a temperature parameter $\lambda$ and Monte carlo samples K. The loss is computed for the sampled architecture, and gradients are calculated with respect to both architecture parameters and network weights using a straight-through estimator. By optimizing these parameters, we aim to find an optimal architecture with minimal parameters.

\section{Experiments and Results}

\subsection{Datasets}
\textbf{FakeAVCeleb}: We evaluate our method on the FakeAVCeleb dataset \cite{khalid2021fakeavceleb}, containing 19,500 fake and 500 real videos of 500 celebrities.\newline
\textbf{SWAN-DF dataset} \cite{korshunov2023vulnerability} is the first publicly available collection of high-quality audio-visual deepfakes, built upon the SWAN database of real-world videos. It contains 24,000 fake and 2,800 real video samples. More details regarding the dataset split is given in the supplementary material.

\subsection{Evaluation methodology}
We employ a two-pronged evaluation strategy. First, we assess model performance on a combined dataset of FakeAVCeleb and SWAN-DF. Second, we evaluate generalization by training on one dataset and testing on the other. Since both databases are biased towards fake videos than the real videos.To mitigate dataset bias, we apply 36 different data augmentation techniques, resulting in 50,742 training, 10,718 validation, and 8,963 test samples similar to \cite{pn2024straight}. Detailed information regarding dataset partitioning and augmentation techniques can be found in the supplementary material.

\subsection{Architecture search and evaluation }
Our experiments involve a two-stage operation selection process. Initially, edges are selected or discarded from a pool $\mathbb{O}^F$. Subsequently, operations from a different pool $\mathbb{O}^S$ are considered. The number of monte carlo samples (K) is varied across different temperature parameter $\lambda$ settings for larger and optimal search space for lower training loss (see training and validation loss graphs for different monte carlo samples and different temperature values in supplementary material). The algorithm stops when the selection of operations within the neural cell stabilizes for both learnable parameters, $\alpha$ and $\gamma$ which is measured as $ E(\alpha)=-\sum_{a,b}\sum_{o \in \mathbb{O}^F} \alpha^o_{(a,b)}\:log(\alpha^o_{ab})$ and $E(\gamma)=-\sum_{a,b}\sum_{o \in \mathbb{O}^S} \alpha^o_{(a,b)}\:log(\gamma^o_{ab})$. Experiments were conducted on V100 Tesla GPUs with 16GB memory. The model was trained using PyTorch with Adam optimizer, a batch size of 8 for 100 epochs, and specific hyperparameters for learning rates, weight decay, and momentum (more details in the supplementary material). The optimal architecture determined through the search process, is trained for 100 epochs with a batch size of 64. Its performance is then evaluated on a held-out test set.

\subsection{Performance comparision}
We assess model performance using standard metrics: Area Under the Curve (AUC) and classification accuracy (ACC) similar to \cite{pn2024straight}.\newline
\textbf{SOTA models:} We compare our method to state-of-the-art audio-visual deepfake detection techniques, including Not made for each other \cite{chugh2020not}, Voice-Face \cite{cheng2023voice}, Audio-Visual Anomaly detection \cite{feng2023self}, ID-Reveal \cite{cozzolino2021id}, Multimodal-trace \cite{raza2023multimodaltrace}, Ensemble learning \cite{hashmi2022multimodal}, and POI-AV \cite{sung2023hearing} and STGS-BMNAS \cite{pn2024straight}.\newline
\textbf{Training:} To ensure fair comparison, all models were trained on our dataset using identical pre-processing steps and adhering to strict data partitioning to prevent overlap between training, validation, and testing sets.






Table \ref{tab:SOTA_cmp} presents a comparison of our proposed GRMC-BMNAS model with SOTA methods using a combined dataset. Our model surpases the recent SOTA models POI-AV \cite{sung2023hearing}, Multimodaltrace \cite{raza2023multimodaltrace}, and ID-Reveal \cite{cozzolino2021id}, STGS-BMNAS \cite{pn2024straight} on combined datasets in terms of accuracy (ACC) and area under the curve (AUC) metrics, while using fewer model parameters. Our model exhibits lower variance compared to STGS-BMNAS, supporting Proposition 1 (Figure given in the supplementary material). Additionally, our model achieves lower mean squared error (MSE) for learnable parameters $\alpha$ and $\gamma$, corroborating Proposition 2 (For figure see supplementary material). Optimal architecture with minimal GPU days is obtained with K=100 and $\lambda=0.1$ is shown in Figure \ref{fig:optimal_arch} . Receiver operating characteristic (ROC) curves for $\lambda=0.1$ and K=100, distinguishing between real and fake data, are provided in the supplementary material.

\textbf{Model performance on unseen data:} Table \ref{tab:gen_data} presents the performance of our model on unseen data. Our model significantly outperforms STGS-BMNAS demonstrating its superior generalization capabilities. This improved performance can be attributed to the model's reduced variance and lower mean squared error as established in previous sections.

\subsection{Ablation study}
\vspace{-0.1cm}
Table \ref{tab:ablation_temp_sample} presents the outcomes of an ablation study examining the influence of temperature and Monte Carlo samples on model performance. Consistent with Proposition 2, increasing the number of Monte Carlo samples generally leads to smaller model sizes and higher AUC values, albeit at the expense of increased computational cost. Based on these findings, an optimal architecture was determined with $\lambda=0.1$ and K=100. Respective architectures produced using different parameter settings can be found in the supplementary material.

\section{Conclusion}

This paper introduces GRMC-BMNAS, a novel architecture search method for audio-visual deepfake detection. Our approach leverages a two-stage Gumbel-Rao Monte Carlo sampling process to efficiently discover optimal architectures. By reducing variance and mean squared error, GRMC-BMNAS surpasses existing methods like STGS-BMNAS in both training efficiency and generalization performance. Extensive experiments are carried out two different data sets such as FakeAVCeleb and SWAN-DF. Both intra and inter dataset experiments are performed to benchmark the performance of the proposed method with other existing AV fake detection techniques. The obtained results indicates the best accuracy and the lower number of hyper parameters for the proposed method.


\end{document}


\title{ Gumbel Rao Monte Carlo based Bi-Modal Neural Architecture Search for Audio-Visual Deepfake Detection\\ \hspace{3.5cm} Supplementary material}


\maketitle
\thispagestyle{empty}

\setcounter{section}{0}
\renewcommand{\thesection}{\Alph{section}}

\section{Gumbel-Rao Monte Carlo estimator}

\begin{figure*}[!h]
  \centering
   \includegraphics[width=0.5\linewidth]{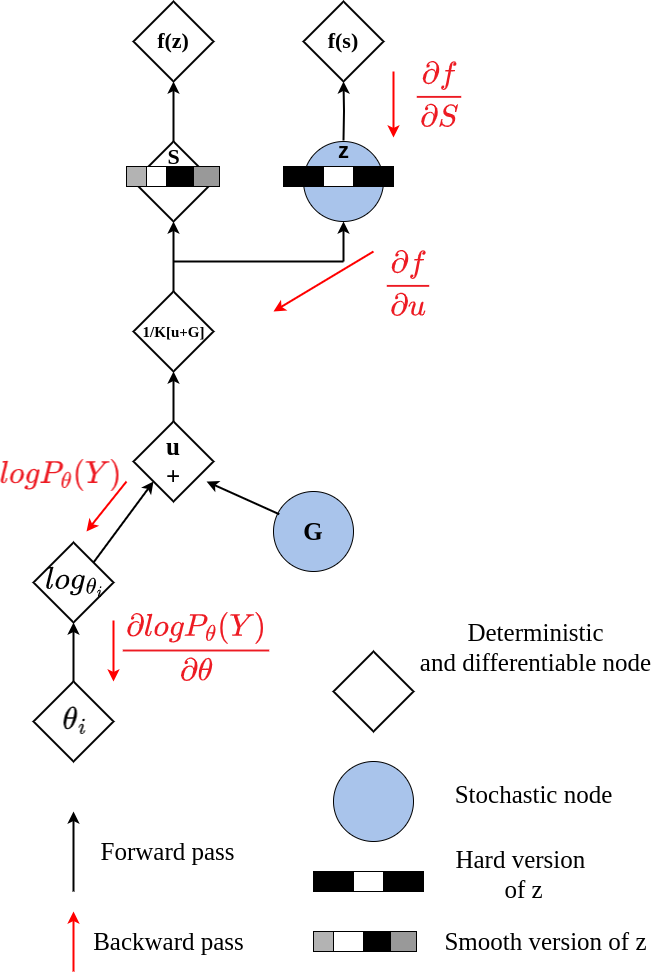}

   \caption{Gradient estimation in GRMC estimator. Black arrows indicates the forward pass and red arrows indicate the backward path}
   \label{fig:grmc_png}
\end{figure*}

Figure \ref{fig:grmc_png} illustrates the process of gradient estimation using the GRMC estimator. During the forward pass, Gumbel sampling is executed, followed by the application of the Rao-Blackwellisation condition, and finally, an average is taken over the Monte Carlo samples. In contrast, the backward pass utilizes the straight-through estimator.

\section{Datasets}
\subsubsection{FakeAVCeleb:} The FakeAVCeleb  built using real videos from VoxCeleb2. VoxCeleb2 is a collection of YouTube videos featuring 6,112 celebrities. It includes more than 1 million videos in the development set and more than 36,000 in the test set, each video containing celebrity interviews. To ensure diversity, videos were carefully selected 500 videos (one per celebrity) from VoxCeleb2, averaging 7.8 seconds each. They maintain the original video dimensions and aim for balanced representation across gender, ethnicity, and age. The database features individuals from four ethnicities (Caucasian, Black, South Asian, East Asian) with 100 videos each, equally divided between males and females. The deepfake dataset, FakeAVCeleb, was created, starting with 500 real celebrities videos. Using various deepfake techniques like face-swapping and reenactment, they generated roughly 20,000 manipulated videos. From these original 500 videos, fake video samples were generated using four of the most recent and advanced deepfake manipulation and synthetic speech generation techniques available at the time. These techniques include Faceswap, Faceswap GAN(FSGAN), Wav2Lip, and Real-Time-VoiceCloning (RTVC). These techniques were applied in combination to create a diverse set of fake video samples. Consequently, the database comprises a total of 20,000 video samples, encompassing both original videos featuring celebrities and synthetic deepfake videos generated using the mentioned techniques.

\subsubsection{SWAN-DF:} SWAN-DF is the first high fidelity publicly available dataset of realistic audio-visual deepfakes, where both faces and voices appear and sound like the target person. The SWAN-DF database is based on the public SWAN database of real videos recorded in HD with the iPhone and iPad Pro in 2019. The original SWAN database comprises 150 subjects, such that 50 data subjects were collected in Norway, 50 data subjects were collected in Switzerland, 5 data subjects in France, and 45 data subjects in India. The age distribution of the data subjects is between 20 and 60 years. The dataset was collected using Apple iPhone6S and Apple iPad Pro (12.9inch iPad Pro). Using a combination of autoencoder-based face-swapping models and blending strategies from DeepFaceLab, the faces of the individuals in each pair were swapped to create deepfake videos. Additionally, voice conversion techniques, also known as voice cloning, were employed to manipulate the audio of the videos. Techniques such as zero-shot YourTTS, DiffVC, HiFiVC, and multiple models from FreeVC were utilized for this purpose.

In total, the SWAN-DF database comprises 24,000 deepfake videos, with each video generated using different combinations of face-swapping models and voice conversion techniques. These videos are available in three different pixel resolutions: 
$160\times160$, $256\times 256$, and $320\times 320$. Alongside the deepfake videos, the database also includes 2,800 real videos, likely serving as the ground truth or reference for comparison purposes.

\subsection{Different Augumentation techniques}
\begin{itemize}
    \item \textbf{RandomRotate:} Rotate video randomly by a random angle within given bounds.
    \item \textbf{RandomResize:} Resize the video by zooming in and out.
    \item \textbf{RandomTranslate:} Shifting video in x and y coordinates.
    \item \textbf{RandomShear:} Shearing video in X and Y directions.
    \item \textbf{CenterCrop:} Extract the center crop of the video.
    \item \textbf{CornerCrop:} Extract the corner crop of the video.
    \item \textbf{RandomCrop:} Extract random crop of the video.
    \item \textbf{HorizontalFlip:} Horizontally flip the video
    \item \textbf{VerticalFlip:} Vertically flip the video
    \item \textbf{GaussianBlur:} Augmenter to blur images using Gaussian kernels.
    \item \textbf{ElasticTransformation:} Augmenter to transform images by moving pixels locally around using displacement fields.
    \item \textbf{PiecewiseAffineTransform:} Augmenter that places a regular grid of points on an image and randomly moves the neighborhood of these point around via affine transformations.
    \item \textbf{Superpixel:} Completely or partially transform images to their superpixel representation.
    \item \textbf{Sequential:} Composes several augmentations together.
    \item \textbf{Add:} Add a value to all pixel intensities in a video.
    \item \textbf{Multiply:} Multiply all pixel intensities with the given value. This augmenter can be used to make images lighter or darker
    \item \textbf{Pepper:} Augmenter that sets a certain fraction of pixel intensities to 0, hence they become black.
    \item \textbf{Salt:} Augmenter that sets a certain fraction of pixel intensities to 255, hence they become white.
\end{itemize}

  
We have applied these 13 different augmentations with different values to make 18 different augmentations except the horizontal flip. Then we horizontally flipped all the 18 augmented videos to get 36 different real augmented videos.
After applying the above-mentioned augmentation techniques for training and validation videos in the
FakeAvCeleb and SWAN-DF datasets we have the split shown in Table \ref{tab:data_aug} and the final train and validation and test split is given by the Table \ref{tab:total_aug_final}. The sample images obtained after augumentation is shown in Figure \ref{fig:eleven_images}.

\subsection{Pre-processing}

\begin{itemize}
    \item \textbf{Video pre-processing:}
    \begin{itemize}
        \item After splitting the database into FakeAvCeleb and SWAN-DF real videos, 36 different augmentations were applied to the real videos in the FakeAvCeleb database, while the SWAN-DF real videos underwent 6 augmentations.
        \item The augmentation applied are salt, pepper, shear, changing intensity, horizontal flip, and many more to balance the database.
        \item Converted all videos of the database to $256\times256$ resolution images with 30fps.
        \item To perform face detection, MTCNN is utilised. MTCNN utilizes
a cascaded framework for joint face detection and alignment.
\item A pre-trained ResNet-34 is utilised to extract the facial image features
    \end{itemize}
    \item \textbf{Audio pre-processing:}
    \begin{itemize}
        \item We extract the audio from video files and and subsequently converting the sampling rate of the audio files to 16,000Hz.
        \item We normalize the values of the audio file by applying mean and convert it to mel spectrogram with window length of 25ms and a shift size of 10ms.
        \item In order to standardize the size of all spectrograms to match the largest one, we duplicate the spectrogram and append it to itself, as the sizes of spectrograms may differ.
        \item We utilise ResNet-34 based pre-trained model to extract $300\times257$ dimensional spectrogram.
    \end{itemize}

\begin{table}[!h]

\centering

\centering
footnotesize
\begin{tabular}{|c|ccc|ccc|}
\hline
 \textbf{Dataset}& \multicolumn{3}{c|}{\textbf{FakeAvCeleb}}                            & \multicolumn{3}{c|}{\textbf{SWAN-DF}}                            \\ \hline
 Split& \multicolumn{1}{c|}{\textbf{Train}} & \multicolumn{1}{c|}{\textbf{Val}} & \textbf{Test}  & \multicolumn{1}{c|}{\textbf{Train}} & \multicolumn{1}{c|}{\textbf{Val}} & \textbf{Test}  \\ \hline
Real & \multicolumn{1}{c|}{13320} & \multicolumn{1}{c|}{1480} & 100 & \multicolumn{1}{c|}{11760} & \multicolumn{1}{c|}{3920} & 560 \\ \hline
 Fake& \multicolumn{1}{c|}{15327} & \multicolumn{1}{c|}{1670} & 4047 & \multicolumn{1}{c|}{10335} & \multicolumn{1}{c|}{3648} & 4256 \\ \hline
\end{tabular}
\caption{Dataset distribution after applying augmentation}
\label{tab:data_aug}

\centering
\begin{tabular}{|c|c|c|c|}
\hline
\textbf{Split} &\textbf{Train}  &\textbf{Val}  &\textbf{Test}  \\ \hline
Real &25080  &5400  &660  \\ \hline
 Fake&25662  &5318  &8303  \\ \hline
Total &50742  &10718  &8963  \\ \hline
\end{tabular}
\caption{Total video count after augmentation}
\label{tab:total_aug_final}

\end{table}

\section{Variance and MSE analysis of STGS and GRMC estimators}

\begin{figure*}[!h]
  \centering
   \includegraphics[width=1\linewidth]{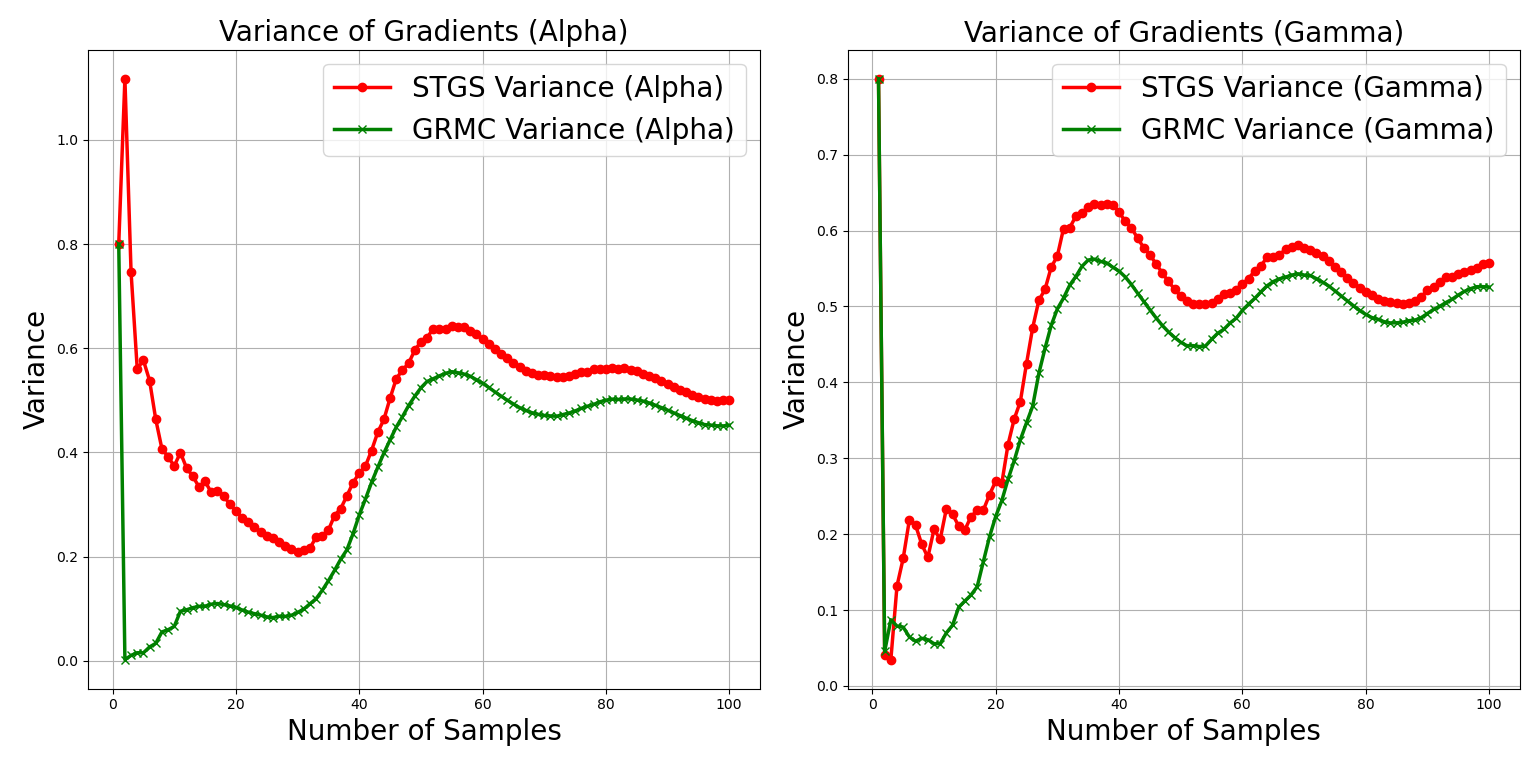}

   \caption{Variance analysis for two learnable parameters $\alpha$ and $\gamma$ for both GRMC and STGS estimators}
   \label{fig:variance_analysis}
\end{figure*}

\begin{figure*}[!h]
  \centering
   \includegraphics[width=1\linewidth]{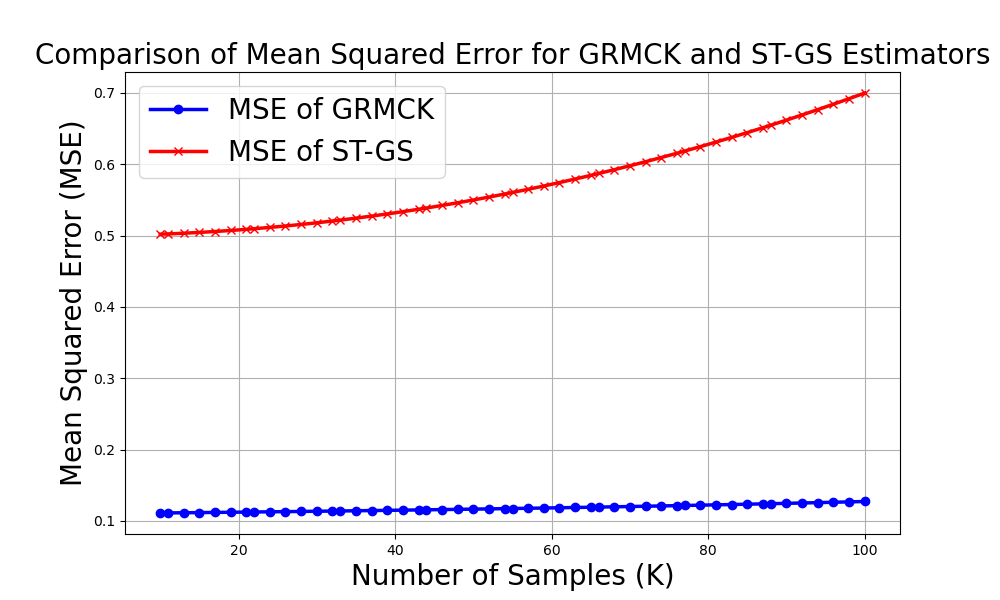}

   \caption{MSE analysis for STGS and GRMC estimators}
   \label{fig:mse_analysis}
\end{figure*}

Figures \ref{fig:variance_analysis} and \ref{fig:mse_analysis} present a comparative analysis of variance and mean squared error (MSE) for STGS and GRMC estimators. These results support propositions 1 and 2, demonstrating the reduced variance and MSE achieved by the GRMC estimator compared to STGS.

    
    

\begin{figure*}[!h]
    \centering
    \begin{subfigure}[b]{0.3\textwidth}
        \centering
        \includegraphics[width=1\textwidth]{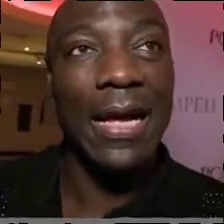}
        \caption*{(a)}
    \end{subfigure}
    \hfill
    \begin{subfigure}[b]{0.3\textwidth}
        \centering
        \includegraphics[width=\textwidth]{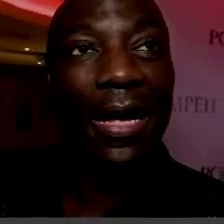}
        \caption*{(b)}
    \end{subfigure}
    \hfill
    \begin{subfigure}[b]{0.3\textwidth}
        \centering
        \includegraphics[width=\textwidth]{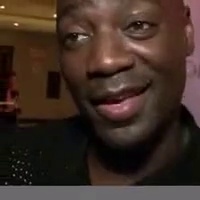}
        \caption*{(c)}
    \end{subfigure}
    
    \begin{subfigure}[b]{0.3\textwidth}
        \centering
        \includegraphics[width=\textwidth]{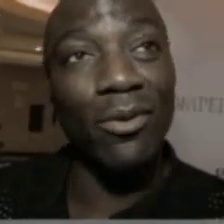}
        \caption*{(d)}
    \end{subfigure}
    \hfill
    \begin{subfigure}[b]{0.3\textwidth}
        \centering
        \includegraphics[width=\textwidth]{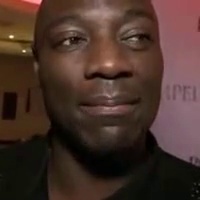}
        \caption*{(e)}
    \end{subfigure}
    \hfill
    \begin{subfigure}[b]{0.3\textwidth}
        \centering
        \includegraphics[width=\textwidth]{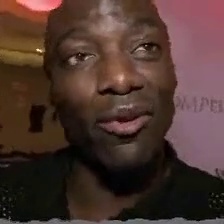}
        \caption*{(f)}
    \end{subfigure}
    
    \caption{Sample images obtained after applying different augmentation techniques: (a) Elastic transformation (b) Add (c) Corner Crop (d) Gaussian blur (e) Center crop (f) Piecewise affine transformation }
    \label{fig:eleven_images}
\end{figure*}

\section{Architectures search}
\subsection{Architectures search on cross dataset generalization}
Figures \ref{fig:arch_fakeav} and \ref{fig:arch_swan} illustrate the generated architectures when trained exclusively on the FakeAVCeleb and SWAN-DF datasets, respectively.
\begin{figure}[htp]
\centering
\begin{minipage}{0.45\textwidth}
\includegraphics[width=\textwidth]{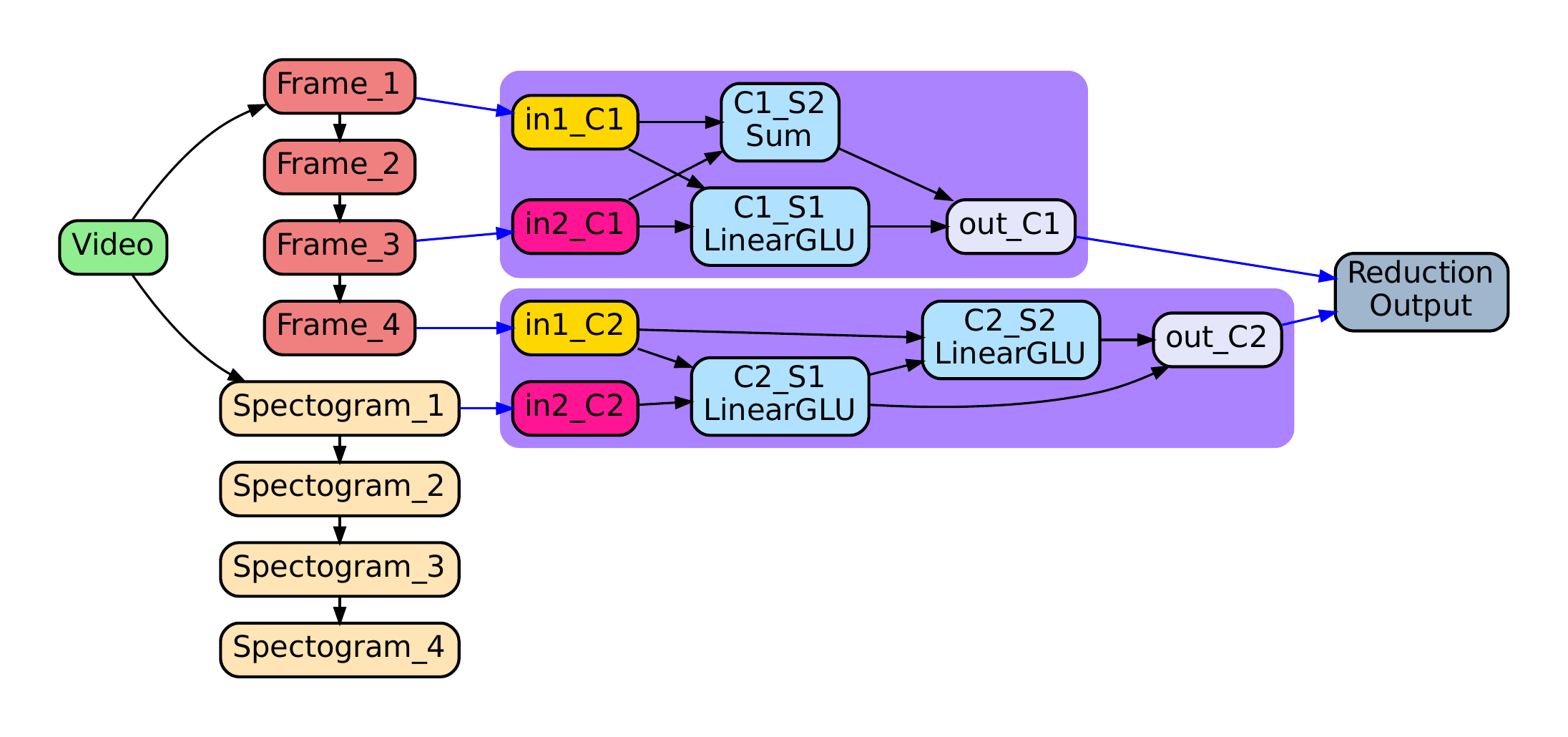}
\caption{Architecture obtained when trained only on FakeAVCeleb database}
\label{fig:arch_fakeav}
\end{minipage}\hfill
\begin{minipage}{0.45\textwidth}
\includegraphics[width=\textwidth]{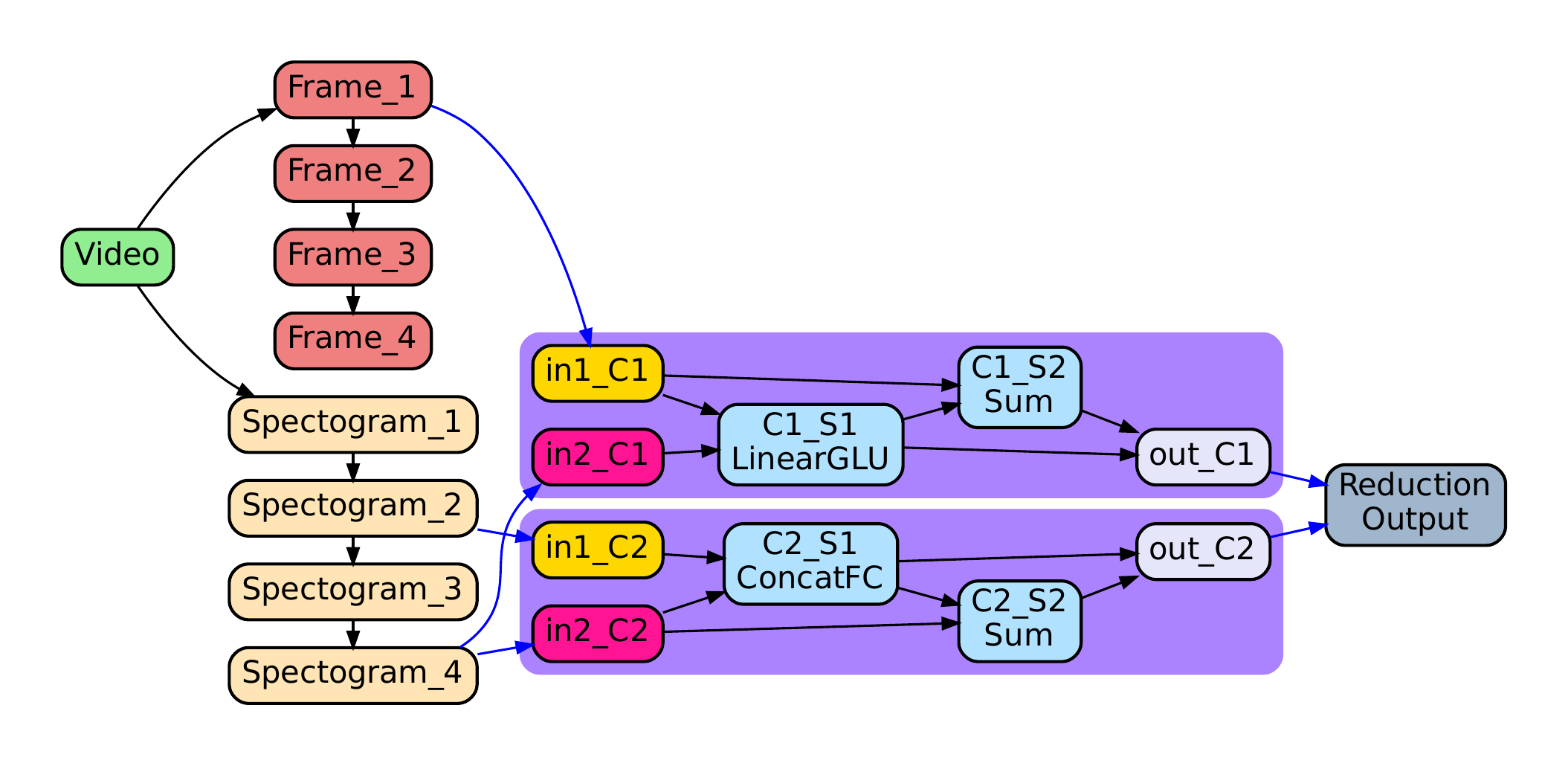}
\caption{Architecture obtained when trained only on SWAN-DF database}
\label{fig:arch_swan}
\end{minipage}\par
\end{figure}

\subsection{Architectures obtained for different temperature and Monte carlo samples}
Figures \ref{fig:arch_temp0.1}, \ref{fig:arch_temp0.5}, and \ref{fig:arch_temp1} display various architectures generated through the ablation study using different temperature and Monte carlo sampling parameters.

\begin{figure}[htp]
\centering
\begin{minipage}{0.45\textwidth}
\includegraphics[width=\textwidth]{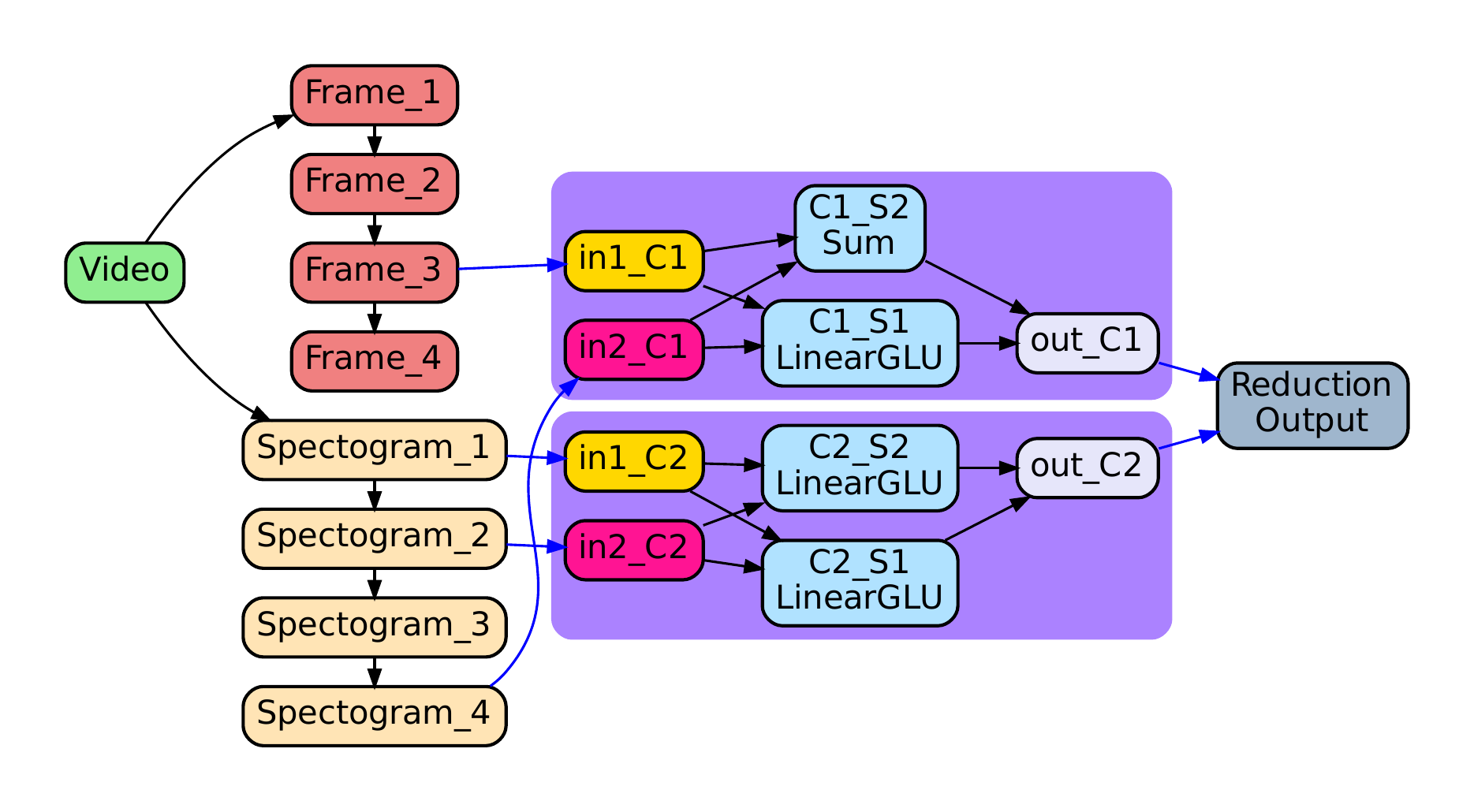}
\caption{Architecture obtained when $\lambda=0.1$ and K=1000}
\label{fig:arch_temp0.1}
\end{minipage}\hfill
\begin{minipage}{0.45\textwidth}
\includegraphics[width=\textwidth]{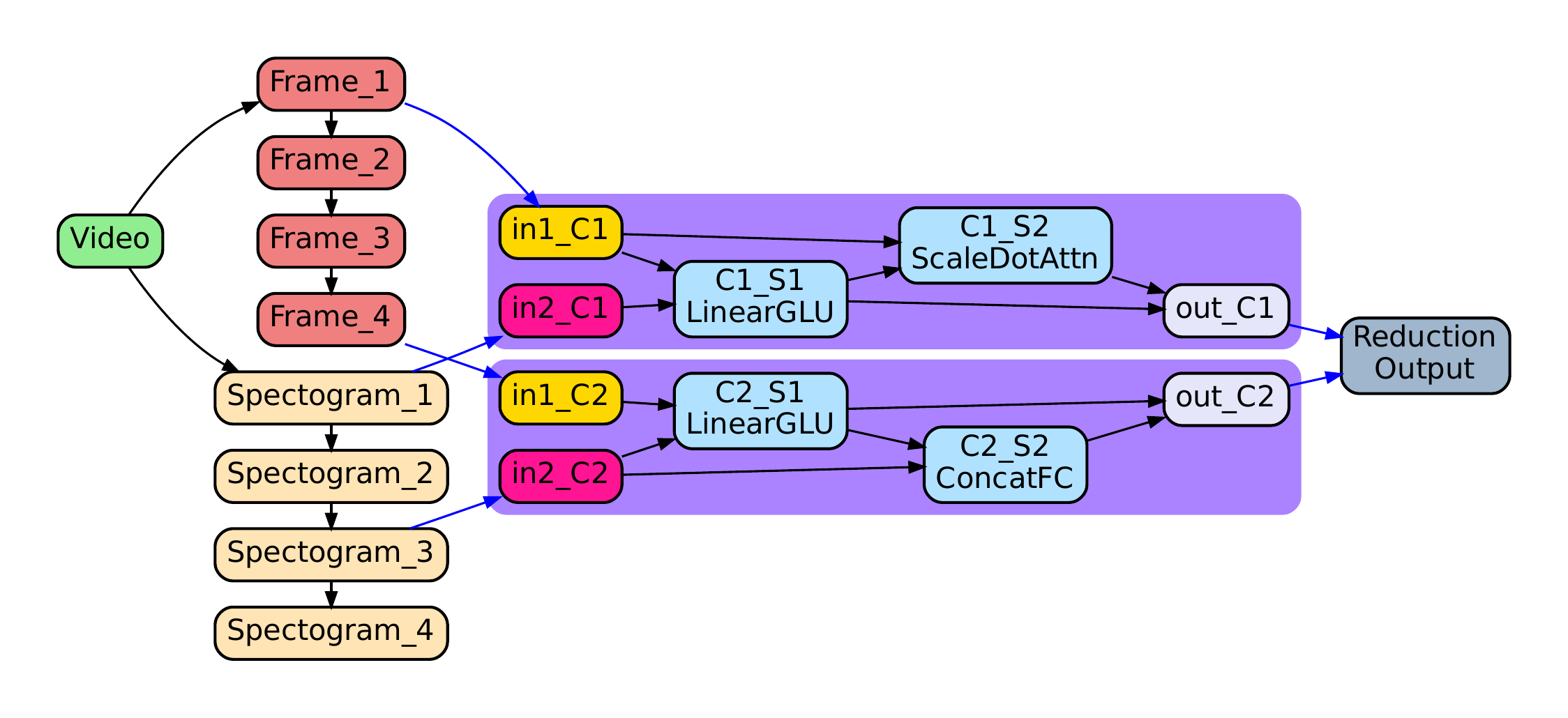}
\caption{Architecture obtained when $\lambda=0.5$ and K=10}
\label{fig:arch_temp0.5}
\end{minipage}\par
\vskip\floatsep
\includegraphics[width=0.45\textwidth]{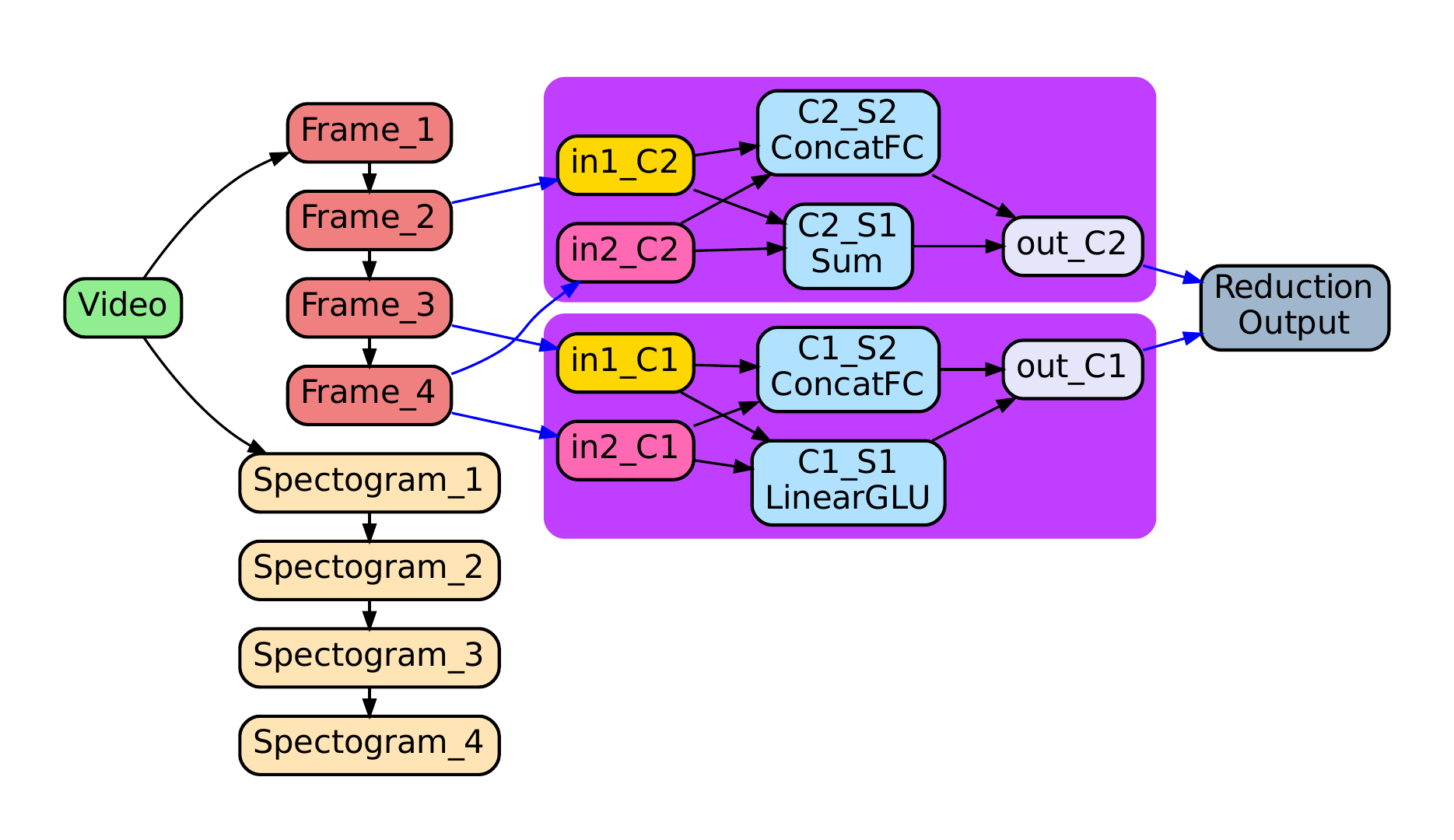}
\caption{Architecture obtained when $\lambda=1.0$ and K=100}
\label{fig:arch_temp1}
\end{figure}

\subsection{ROC curves for real and fake data}
Figures \ref{fig:roc_real} and \ref{fig:roc_fake} illustrate the ROC curves for Softmax, STGS-BMNAS, and GRMC-BMNAS. The GRMC-BMNAS consistently outperforms the other methods in terms of roc for both real and fake classes.

\begin{figure}[htp]
\centering
\begin{minipage}{0.45\textwidth}
\includegraphics[width=\textwidth]{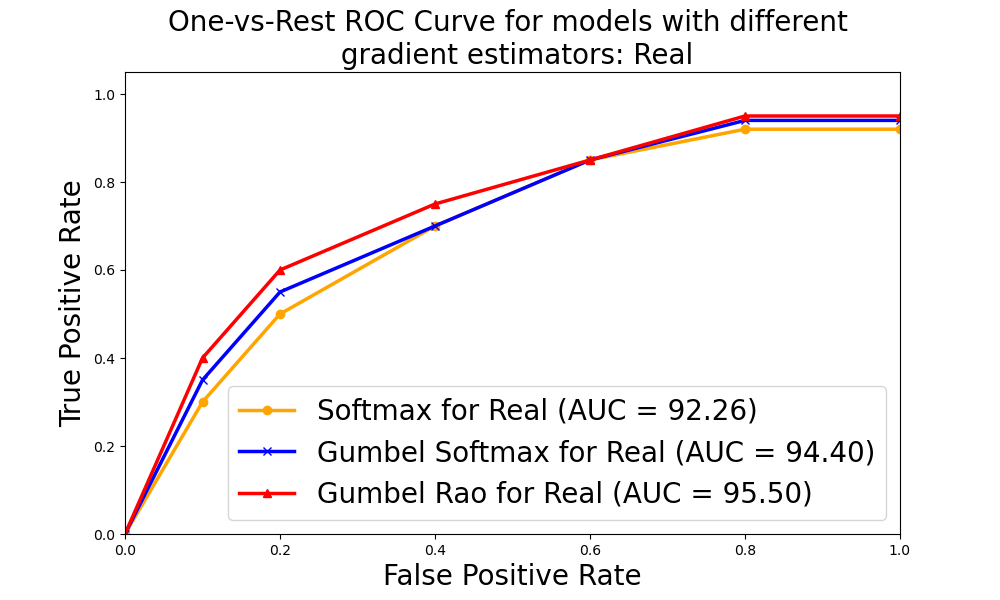}
\caption{ROC curve comparison for the real class: Softmax, STGS-BMNAS, and GRMC-BMNAS.}
\label{fig:roc_real}
\end{minipage}\hfill
\begin{minipage}{0.45\textwidth}
\includegraphics[width=\textwidth]{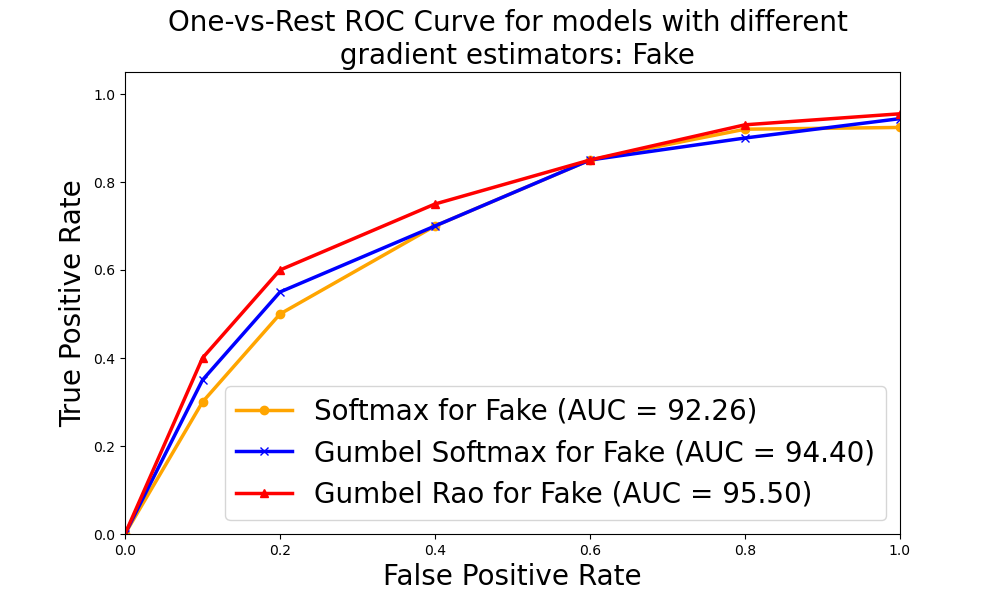}
\caption{ROC curve comparison for the fake class: Softmax, STGS-BMNAS, and GRMC-BMNAS}
\label{fig:roc_fake}
\end{minipage}\par
\end{figure}

\subsection{Stable Variance and Improved Performance of GRMC Estimator at Low Temperatures}
\begin{itemize}
    \item \textbf{Improvement below t=0.1:} The trend from the ablation study indicates that the performance of the GRMC estimator improves as the temperature $\lambda$ decreases below 0.1.
    \item \textbf{Variance Behavior:}For the Gumbel Softmax (GS) and Straight-Through Gumbel Softmax (GS-ST) estimators, the variance grows unbounded as the temperature 
t decreases, following an asymptote of $\mathcal{O}(\frac{1}{\lambda})$ approaches zero, the variance increases significantly.
\item In contrast, for the GRMC estimator, the variance does not grow unbounded as $\lambda$ decreases. This suggests that the GRMC estimator has a more stable variance behavior at low temperatures, making it potentially more robust in such conditions.
\end{itemize}

\subsection{Failure cases}
\begin{figure*}[!h]
  \centering
   \includegraphics[width=1\linewidth]{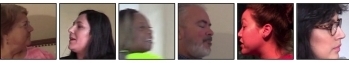}

   \caption{Failure cases for the proposed approach. All the videos comes from FakeAVCeleb database}
   \label{fig:falure_cases}
\end{figure*}

While no failure cases were observed in the frontal video dataset SWAN-DF, some failures were identified in the FakeAVCeleb database. These instances are highlighted in Figure \ref{fig:falure_cases}.

\section{Limitations}
Our current method employs a manual tuning of temperature and Monte Carlo samples to balance exploration and exploitation. To enhance this, we propose adaptive strategies. First, we advocate for dynamically adjusting the number of Monte Carlo samples based on the estimated variance of the current solutions, prioritizing efficiency. Second, we suggest replacing the static temperature with a tempered Gumbel softmax approach, allowing for a more nuanced exploration-exploitation balance. This adaptive approach is expected to improve the search efficiency and solution quality.